\documentstyle[aas2pp4,flushrt,psfig]{article}
\def\simless{\mathbin{\lower 3pt\hbox
   {$\rlap{\raise 5pt\hbox{$\char'074$}}\mathchar"7218$}}} 
\def\simgreat{\mathbin{\lower 3pt\hbox
   {$\rlap{\raise 5pt\hbox{$\char'076$}}\mathchar"7218$}}}

\begin{document}

\title{Annual modulation in IDV of quasar 0917+624 due to Interstellar Scintillation}

\author{\sc B.\ J.\ Rickett\altaffilmark{1}}
\affil{Department of Electrical \& Computer Engineering \\
University of California, San Diego, La Jolla, CA 92093-0407}
\author{\sc A. Witzel\altaffilmark{2}, 
A. Kraus\altaffilmark{2}, 
T.P. Krichbaum\altaffilmark{2}}
\affil{Max-Planck-Institut f\"ur Radioastronomie,\\
Auf dem H\"ugel 69, D-53121 Bonn, Germany }
\author{S.J. Qian\altaffilmark{3}}
\affil{Max-Planck-Institut f\"ur Radioastronomie,
Auf dem H\"ugel 69, D-53121 Bonn, Germany\\
and Beijing Astronomical Observatory, National Astronomical Observatories (CAS), Beijing 100012, China}
\altaffiltext{1}{e-mail: bjrickett@ucsd.edu}
\altaffiltext{2}{e-mail: awitzel, akraus tkrichbaum \& sqian @mpifr-bonn.mpg.de}
\altaffiltext{3}{e-mail:rqsj@class1.bao.ac.cn}

Short title:  Annual Modulation in IDV 0917+62 due to ISS

\begin{abstract}

The quasar 0917+624 has been one of the best studied intraday variable (IDV) radio
sources. However, debate continues as to whether the underlying cause
is intrinsic or extrinsic. Much previous work has assumed the IDV 
to be intrinsic and which implies an extraordinarily compact source for
the radio emission; in contrast, an extrinsic variation due to 
interstellar scintillation (ISS) implies a relatively larger
source diameter, though at the smaller end of the range expected for 
relativistic jet models. Kraus et al. 
(1999) 
reported a marked slowing 
of the IDV at 6cm wavelength in September 1998,
and suggested a change in the source was responsible.
However, here we show that the slowing is consistent with the annual
modulation in scintillation time-scale 
expected for ISS, under the assumption that the scattering medium
moves with the local standard of rest (LSR).  The ISS
time scale is governed by the ISS spatial scale 
divided by the Earth's velocity relative to the scattering plasma.
It happens that in the direction 
of 0917+624 the transverse velocity of the Earth with respect 
to the LSR varies widely with a deep minimum in
the months of September to November. Hence the slowing of the 
IDV in September 1998 
strongly suggests 
that ISS, rather than intrinsic 
variation of the source is the dominant cause of the IDV. 

Subject headings: Scattering---ISM: kinematics and dynamics---
quasars:individual:B0917+624---Radio continuum:general

\end{abstract}


\section{INTRODUCTION}

Rapid (faster than a day) variations in the radio flux density of some 
extra-galactic sources have attracted considerable attention
in the years since they were first reported by Heeschen et al.\ (1987).
Via the classic light travel time argument for intrinsic 
variations, the short variability time 
scales have been used to infer an extremely small
physical size of the radio emitting regions;
the associated apparent brightness temperatures
are very much higher than expected from 
relativistic jets from active galactic nuclei (AGNs) 
emitting Doppler-boosted radio beams (Qian et al., 1991).  
The canonical jet model successfully explains 
``superluminal'' motion in VLBI observations, invoking 
Lorentz-factors (bulk jet speeds)
in the range of 1-20. In order to also explain the high brightness
temperatures infered from IDV one needs very small
viewing angles and hence such jets can explain brightness 
temperatures of up to $10^{16}$ K.   An intrinsic view of the
most extreme IDV at cm-wavelengths
would imply imply brightness temperatures up to
$10^{21}$ K (e.g. Kedziora-Chudczer et al. 1997 and
Dennett-Thorpe \& de Bruyn, 2000a). This stretches the 
jet model too far and in both  cases the authors
proposed ISS rather than intrinsic interpretations.

Quirrenbach et al.\ (1992) presented the results of daily monitoring 
of a complete sample of flat spectrum sources at 6cm and 
11cm. They found a low level of IDV in most of the 
sources observed, which also had compact VLBI structure.  
They concluded that this low level of variabilty seen in almost 
all such sources was ISS. However, in addition they argued
that about 25\% of flat spectrum
sources also showed intrinsic variations that can reach
amplitudes as high as 20\%.  0917+624 was one
of these strongly variable sources, for which
intrinsic models have been invoked 
(Wagner \& Witzel, 1995 and references therein). 

On the other hand, Rickett et al. (1995) [R95]
demonstrated that ISS could successfully account for 
the IDV in 0917+624 with a source brightness of about
$6 \times 10^{12}$K, by assuming that the path length for the 
scattering medium was reduced to 200 pc, from the 930 pc predicted
from the Taylor \& Cordes (1993) model for the distribution of 
electrons in the Galaxy [TC93].  In R95 no consideration was
given to the influence of the change in the Earth's velocity relative to the
scattering medium. In this paper we reconsider the accumulated 
observations of the source at 6cm by the Bonn group and study
how the time scale and modulation index 
(rms normalized by mean flux density)
vary over the year and compare with predictions.  

The 0917+624 observations of September 1998 showed 
a remarkably different
character -- with the flux varying slowly and nearly
linearly over the course of the 5-day observing span.
This was reported by Kraus et al.\ (1999) [K99], who
suggested that there had been a pronounced change
in the emitting properties of the source, which then 
had recovered by February 1999 to explain the return to
its normal IDV behavior observed at that time.
We conclude that this anomolous behavior
supports the ISS interpretation since it is simply
explained by the changing Earth velocity.

\section{Observations and Data Analysis}

We have collected together all of the observations of 0917+624
at 6cm since 1989 made at Effelsberg and at the VLA 
in flux density monitoring campaigns lasting more than 24 hours. 
We choose 6cm as the wavelength most commonly observed 
(since it has good signal-to-noise ratio).  

We have used a structure function analysis on 
each time series to estimate the characteristic time scale 
and the rms amplitude, as used by R95.  The sampling interval
for the flux measurements was typically somewhat 
variable, so the structure function was binned
into uniform intervals ($\delta \tau$) in time lag 
about equal to the average time between samples
(typical values were 0.05 to 0.1 day).
The maximum time lag computed is about half the duration
of each data set. The structure function $D(\tau)$ 
is first corrected for the noise level, 
which is estimated as $D$ at the smallest 
time lag and subtracted from $D(\tau)$ at each
lag.  A region in time lag is then identified where
$D(\tau)$ saturates. However, it should be noted that, 
since  $D$ is estimated from a finite sample of a 
stochastic variation, it will fluctuate about a saturation
value rather than tending to an asymptote.
The flux variance is estimated as half of the 
average of $D$ over the saturation region;  the time scale is estimated
as the lag where $D$ crosses half the saturation level.
Table 1 shows the results for each observation with 
estimated 
rms 
errors, which are dominated by the
estimation error from the modest number of independent 
samples in each series.

\begin{table*}[htb]
\tablecaption{0917+624 IDV parameters at 6 cm}
\begin{tabular}{ccccccc}
\tableline
\tableline
Date & day no. & Duration &  $\tau$ & $<S>$ & $S_{\rm rms}$ & Modulation index  \\
(yr/m/d) &  & (days) & (days)  & (Jy) & (Jy) &  \\
\tableline
88/4/3  &  94 &  4.4 &  $0.22 \pm.10$ & 0.71 & 0.025 & 0.035 \\
88/6/17 & 169 &  4.6 &  $0.33 \pm.15$ & 1.25 & 0.068 & 0.054 \\     
88/12/30 & 365 & 6.1 &  $0.28 \pm.12$ & 1.44 & 0.086 & 0.059 \\
89/5/6  & 126 &  5.3 &  $0.13 \pm.05$ & 1.45 & 0.060 & 0.042  \\
89/5/6 &  126 &  4.9 &  $0.19 \pm.07$ & 1.51 & 0.060 & 0.040  \\
89/12/26 &  360 &  5.9 &  $0.34 \pm.10$ & 1.50 & 0.083 & 0.056  \\
90/2/11 &  42 & 25.1 &  $0.17 \pm.02$ & 1.46 & 0.049 & 0.034  \\
91/12/31 & 365 & 6.9 &  $0.29 \pm.14$ & 1.19 & 0.030 & 0.025  \\
93/4/12 & 102 &  2.3 &  $\simgreat 0.13 (-.08)$ & 1.52 & 0.018 & 0.012  \\
93/6/19 & 170 &  2.6 &  $\simgreat 0.15 (-.07)$ & 1.59 & 0.050 & 0.032  \\
97/12/6 & 340 &  2.7 &  $\simgreat 0.55 (-.25)$ &  1.42 & 0.069 & 0.049  \\
97/12/28 & 362 & 5.4 &  $0.28 \pm.14$ & 1.45 & 0.074 & 0.051 \\
98/9/19 & 262 &  5.0 &  $\simgreat 1.33 (-.9)$ &  1.51 & $\simgreat 0.029$ & $\simgreat 0.019$  \\
99/2/9  &  40 &  2.2 &  $\simgreat 0.37 (-.2)$ &  1.53 & 0.089 & 0.058  \\

\tableline
\end{tabular}
\end{table*}

The procedure works well provided that the structure
function does indeed start to saturate before
the maximum lag, as is seen for nearly all of
the observations.  However, in  September 1998 
the flux did not show several maxima and 
minima per day as it normally does, rather
there was an approximately linear increase
over 4.5 days.  The associated structure
function does not show signs of saturation (K99);
so here we set a lower bound on the time
scale and on the flux variance by treating $D$ 
at the maximum lag as if it were saturated.
We consider this data set as if it were a sample
of a stochastic variation with a time scale longer 
than its duration and obtain a substantially longer
time scale than from the rest of the observations.

\section{Annual Modulation in ISS Parameters}

\begin{figure}[tbh]
\psfig{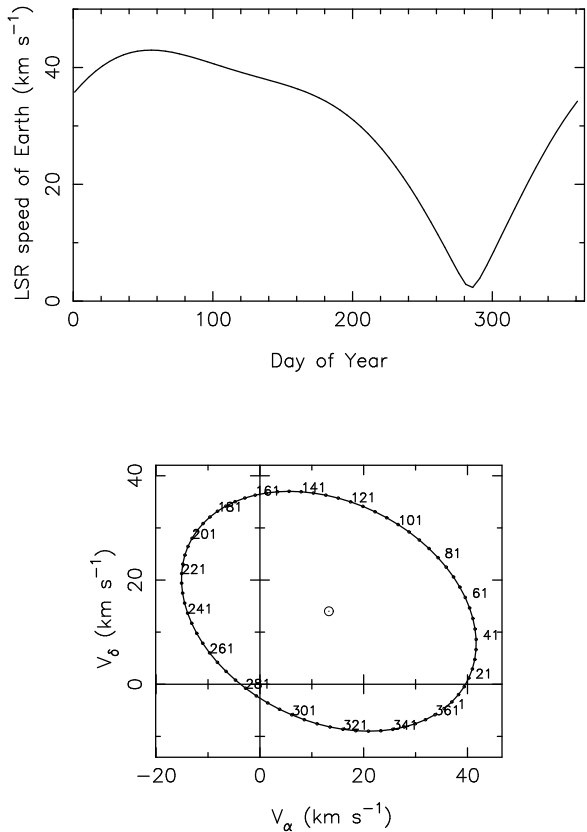}
\caption{Earth velocity (in km/s) relative to the LSR projected
transverse to the direction toward 0917+624. 
a) Upper panel: speed versus day of year.  b) Lower panel: velocity plotted as 
$V_{\alpha}$ and $V_{\delta}$ at 5-day intervals.
Note the time near day 285 where the velocity is close to zero.}
\end{figure}

The time scale for ISS is determined by the ratio
of two quantities -- the spatial scale of the scintillation pattern
and the relative speed of the Earth through the pattern.
The pattern scale depends on the line-of-sight distribution
and the wavenumber spectrum of the
scattering material and also on the angular structure
of the compact component in the radio source.
The ISS theory and model used by R95 assumed 
the source to be circularly symmetric and sufficiently extended 
to partially quench the scintillations, in which case the same theory
can be applied for both weak and strong refractive scintillation;
further it assumes the scattering plasma to be
isotropic and to be in a thick 
layer with a Gaussian profile.  The natural
scale height to choose is that of the ionized 
disk component in the TC93 model, which gives
a characteristic path length of 930 pc for the 
density variance at the latitude of 0917+624.  
However, as noted above a smaller path length 
(200 pc) was found necessary by R95 
to fit the observations with a speed that was (arbitrarily) 
assumed to be 50 km/s.  In this section we use the
same R95 model for the spatial scale and calculate the 
Earth's transverse velocity explicitly versus day 
of year, under the assumption that
the entire scattering medium moves with the LSR and then
compare the observed and predicted time scales.
This assumption is consistent with ISS observations
of the pulsar B0809+74  by Rickett, Coles
and Markkanen (2000), who concluded that
the velocity of the scattering medium is
within 10 km/s of the LSR.

Figure 1 shows the Earth's velocity relative to the LSR
projected perpendicular to the line of sight to 0917+624
versus day of year.  The velocity due to the Earth's 
orbital motion around the Sun happens to nearly 
cancel the projected velocity of the Sun relative to the LSR
near day number 285.  Thus the predicted
ISS speed shows a large annual modulation from a maximum 
of 40 km/s to a minimum of 4 km/s.  
In Figure 2a we overplot the observed and predicted 
time scales against the day of year,
including as thin lines the bounds derived from predictions
for a range of offsets in plasma velocity relative to the LSR 
covering $\pm 5$ km/s in the right ascension
direction and $\pm 5$ km/s in the declination direction.

\begin{figure}[tbh]
\psfig{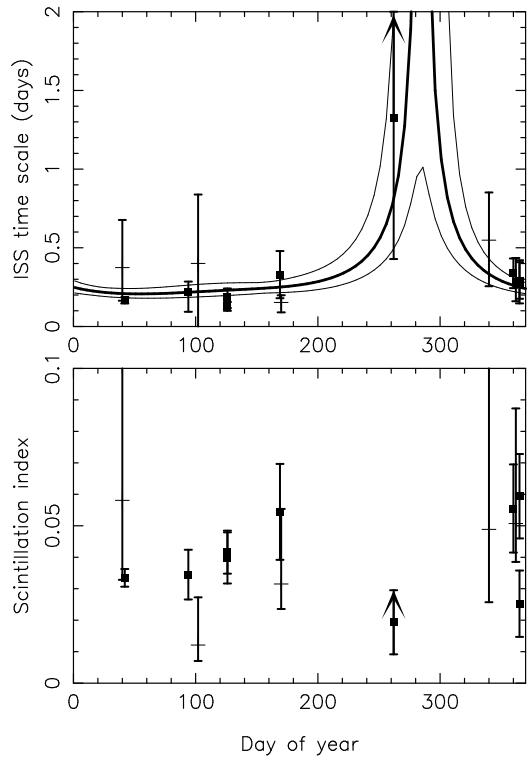}
\caption{(a) Upper panel: Time scale versus day of year for 
the 6 cm observations of 0917+624. Thick curve is 
ISS prediction for scattering plasma 
moving with the Local Standard of Rest; thin curves are 
upper and lower bounds obtained for velocities ranging over
$\pm$5 km/s in RA and in Dec relative to the LSR. 
(b) Lower panel: Apparent scintillation index versus day of year.
Observations shorter than 3 days are plotted as horizontal bars 
longer observations as a filled squares. }
\end{figure}

The striking feature of the predicted time scale is that
it should remain fairly constant over much of the year
and exhibit a substantial increase over days 250-330
(September--November).  Even though there have been 13 
independent observations of the 6cm IDV, it happens that only
one of them lies in this period of predicted slow ISS.
Thus the apparent constancy of IDV behavior
is readily understood as is the one anomolous
observation.  Overall we find a general agreement between the 
observations and the predictions for the time scale.

Turning to the scintillation (or modulation) index,
there is a question of what should be the normalizing
flux, since the source may has some of its
flux in components of too large a diameter to
scintillate ($>1$ mas). The VLBI maps of Standke et al.\ (1996) and 
new unpublished maps (Krichbaum, priv. comm.)
show that the source exhibits a core jet structure composed 
out of a number of unresolved ($<1$ mas) and partially resolved 
($>1$ mas) components.
However, in considering any annual change in 
scintillation index this normalizing uncertainty
is unimportant. What matters is whether there
should be any annual change in the ISS variance.
Since the ISS variance is given by the integral
of the intensity power spectrum over two-dimensional 
wavenumber space (Little \& Hewish, 1966), 
the result is independent of the
orientation of the velocity vector, 
even if the wavenumber spectrum is non-circular
due to any anisotropy in the irregularity
spectrum and/or anisotropy in the source visibility function.
Thus there should be no annual change in the
scintillation index. Figure 2b shows this as
the observed scintillation index versus day of year.
The  fact that the observations do not show
an obvious annual variation is consistent
with this model, 
and though there are some discrepant values
they do not appear to be statistically significant.

It has been argued that the interstellar plasma irregularities
are commonly anisotropic (e.g. Desai et al.\ 1994).  
The orientation of the velocity
relative to the major axis of such anisotropy will vary
during a year, as the direction of the Earth's motion 
changes. In such a case the spatial scale of the ISS
pattern will show 
a variation over six months, 
which can combine
with the changing Earth velocity to create a variety
in the predicted annual time scale curves.  If in addition the source
structure is not circular its angular orientation
would also affect the spatial scale and so will 
contribute to the annual modulation. 
(Dennett-Thorpe and de Bruyn, 2000a and 2000b).

\section{Discussion and Conclusions}

It is clear from Figure 2 that ISS governs the time scale 
of the IDV,
unless there was a transient slowing of intrinsic variability that
coincidentally echoed the behaviour expected for ISS.
Hence we conclude that the IDV of 0917+624 at 6cm is  
predominantly
caused by scintillation in the interstellar medium. 
If simultaneous intrinsic variations
faster than 6 hours
are also present their rms amplitude must be 
less than 1\% (obtained from the structure function analysis
of the data from September 1998).

This confirms the work of R95 and also
opens up the possibility of using these seasonal ISS effects to
explore anisotropy in both the source and the scattering plasma.
Our conclusion does not at all reduce the importance
of intrinsic variations in explaining source variations on times of
months to years, as particularly evident in VLBI maps
and longer term flux changes.  Indeed both phenomena
are needed for a full understanding of AGN behaviour.

As discussed by R95 the time scale observations jointly constrain 
the scattering distance and source diameter in a fashion that depends 
on the distribution of the scattering plasma.
The sense of this constraint is that a large
distance implies a smaller source diameter and vice versa.
In the R95 model the scattering distance was
about 200 pc, in preference over the  930 pc expected under the
TC93 model. The accompanying
source diameter at 6cm was about 0.07 mas.
Models with a larger scale height required a smaller
source size and vice versa. R95 assumed that the velocity 
was fixed at 50 km/s.  Now with a better model for the velocity 
over the course of a year we can re-examine the distance/diameter 
constraint.  The best estimate of the ISS time scale
comes from the 25-day observations from February 1990
(Quirrenbach et al. 2000).
Since the LSR velocity was 42 km/s at that time (see Figure 1), 
the nominal 200 pc medium thickness is only reduced 
by the ratio 42/50 to 170 pc. 
In a related paper
(Rickett and Lyne, in preparation, 2000) we consider this constraint in 
the light of diffractive ISS observations of 
pulsar B0917+63, which at a nominal distance of 760 pc
probes the same path through the Galaxy.

The agreement in time-scale with the prediction based
on assuming that the medium is stationary in the LSR
allows us to put an upper bound on the range of velocities 
that contribute to the scattering. If the distribution of 
speeds in the scattering medium is a Gaussian function, 
its standard deviation must be no more than the
slowest effective ISS speed, (defined by the
spatial scale in the model divided by the observed time scale).  
This bounds the standard deviation
in speed at about 8 km/s.  
In comparison transverse differential Galactic rotation amounts to
about 1km/s at 200 pc along this line of sight.
Though there is no distinction 
possible between a spread of velocities due to shear or 
due to turbulence in the medium, this result puts an upper limit
of 8 km/s for any turbulent velocities in the scattering plasma.
This is one of the few formal constraints on turbulent plasma 
velocities that have been extracted from three decades of ISS work.
Future observations sampling the full annual cycle of ISS 
will help determine the velocity distribution in the ISM,
though the influence of anisotropic scattering will also have to be modelled.

Evidently, there is a strong incentive to verify the annual 
variation and to study whether similar phenomena exist for other
IDV sources, several of which are in 
directions for which a substantial annual modulation should be
expected.  Preliminary examination for IDV sources 
0716+714 and 0954+658 do not show the predicted annual
modulation in IDV time scale.  Thus the present result for 0917+624
does not rule out the existence of intrinsic IDV in these 
sources.  A program of IDV observations of 0917+624 has been 
started in September 2000 and is continuing in order
to verify 
the slow ISS phenomenon for 0917+624.
During the final preparation of this paper, we have received a draft of
a paper by Jauncey \& Macquart, in which they independently report the
same annual modulation and its ISS interpretation, derived 
from 0917+624 data in the literature.

Acknowledgement.  We thank the many people involved in helping to make these
observations over many years, both at Effelsberg and the VLA.
BJR acknowledges support from the NSF under grant AST-9988398.


\begin{thebibliography}{}

\bibitem[]{} 
Dennett-Thorpe, J. \& de Bruyn, A.G., 2000a, \apj, 529, 65

\bibitem[]{} 
Dennett-Thorpe, J. \& de Bruyn, A.G., 2000b, Proceedings of IAU Colloquium 182, Guiyang, China, April, 2000

\bibitem[]{} 
Desai, K.M., Gwinn, C.R. \& Diamond, P.J., 1994, Nature, 372, 754

\bibitem[]{}
Heeschen, D. S., Krichbaum, T.P., Schalinski, C.J. \& Witzel, A.,
1987, \aj, 94, 1493

\bibitem[]{} 
Kedziora-Chudczer, L., Jauncey, D.L., Wieringa, M.H.,
Walker, M.A., Nicolson, G.D., Reynolds, J.E. and Tzioumis, A.K., 1997,
\apj,  490,  L9

\bibitem[]{}
Kraus, A., Witzel, A.,  Krichbaum, T. P., Lobanov, A.P., Peng, B. and Ros, E.,
1999, \aap, 352, L107 [K99]

\bibitem[]{}
Little, L.T. \& Hewish, A., 1966, \mnras, 134, 221

\bibitem[]{}
Qian, S.J., Quirrenbach A., Witzel, A.,  Krichbaum, T., Hummel, C.A. \&
Zensus, J.A., 1991, \aap, 241, 15

\bibitem[]{}
Quirrenbach A., Witzel, A.,  Krichbaum, T., Hummel, C.A., Wegner, R.,
Schalinski, C.J., Ott, M., Alberdi, A. and Rioja, M., 1992, 
\aap, 258, 279

\bibitem[]{}
Quirrenbach, A., Kraus, A., Witzel, A., Zensus, J.A., Peng, B., Risse, M.,
Krichbaum, T.P., Wegner, R., Naundorf, C.E., 2000,  \aaps, 141, 221.

\bibitem[]{} 
Rickett, B.J., Quirrenbach, A., Wegner, R., Krichbaum, T.P. 
 \&  Witzel, A., 1995, \aap, 293, 479 [R95]

\bibitem[ ]{ }
Rickett, B. J., Coles, W. A., \&  Markkanen, J. 2000, \apj, 533, 304

\bibitem[]{} 
Standke, K.J., Quirrenbach, A., Krichbaum, T.P., et al.\, 1996, \aap, 306, 27

\bibitem[]{}
Taylor, J.H. and Cordes, J.M., 1993, \apj, 411, 674 [TC93]


\bibitem[]{}
Wagner, S.J.  \& Witzel, A., 1995, \araa, 33, 163

\end{thebibliography}
\end{document}